\documentclass[runningheads]{llncs}
\usepackage[utf8]{inputenc}

\usepackage{hyperref}

\usepackage{csquotes}

\usepackage{graphicx}
\graphicspath{{images/}}

\usepackage{xcolor}

\usepackage{multirow}

\usepackage{makecell}

\usepackage{array}
\usepackage{float}

\usepackage{enumitem}

\begin{document}

\title{Which RESTful API Design Rules Are Important and How Do They Improve Software Quality? A Delphi Study with Industry Experts}


%
\titlerunning{Which RESTful API Design Rules are Important?}
%
\author{
    Sebastian Kotstein\inst{1}\orcidID{0000-0001-7113-4450} \and
	Justus Bogner\inst{2}\orcidID{0000-0001-5788-0991}
}
\authorrunning{S. Kotstein and J. Bogner}
%
\institute{
	Reutlingen University, Herman Hollerith Zentrum, Germany\\
	\email{sebastian.kotstein@reutlingen-university.de}
	\and
	University of Stuttgart, Institute of Software Engineering, Germany\\
	\email{justus.bogner@iste.uni-stuttgart.de}
}
\maketitle              
\begin{abstract}
Several studies analyzed existing Web APIs against the constraints of REST to estimate the degree of REST compliance among state-of-the-art APIs. These studies revealed that only a small number of Web APIs are truly RESTful. Moreover, identified mismatches between theoretical REST concepts and practical implementations lead us to believe that practitioners perceive many rules and best practices aligned with these REST concepts differently in terms of their importance and impact on software quality. We therefore conducted a Delphi study in which we confronted eight Web API experts from industry with a catalog of 82 REST API design rules. For each rule, we let them rate its importance and software quality impact. As consensus, our experts rated 28 rules with high, 17 with medium, and 37 with low importance. Moreover, they perceived usability, maintainability, and compatibility as the most impacted quality attributes. The detailed analysis revealed that the experts saw rules for reaching Richardson maturity level 2 as critical, while reaching level 3 was less important. As the acquired consensus data may serve as valuable input for designing a tool-supported approach for the automatic quality evaluation of RESTful APIs, we briefly discuss requirements for such an approach and comment on the applicability of the most important rules.
\end{abstract}

\section{Introduction}
\label{sec:introduction}
Around 2000, Roy T. Fielding formalized the idea of \enquote{how a well-designed Web application behaves} by introducing a novel architectural style that he named \enquote{Representational State Transfer} (REST)~\cite{thesis:Rest2000}.
In detail, he defined REST as a set of constraints that a Web application must fulfill to improve several quality attributes like scalability, efficiency, and reliability.
The motivation for this was the concern that the Web as an open platform, i.e. not being under the control of a central entity, would not scale without rules and constraints to govern interactions and information exchanges~\cite{thesis:Rest2000,book:RestApiDesignRulebook2011}.

Although REST mainly relies on well-established Web standards like Uniform Resource Identifiers (URIs) and is aligned with the features of the Hypertext Transfer Protocol (HTTP), REST and its constraints are not an official standard~\cite{article:Rodriguez2016}.
Moreover, Fielding's initial formulation of REST and its constraints describe the recommended behavior of a Web application from a high-level perspective instead of providing developers with detailed instructions on how to achieve this behavior.
As a consequence, REST leaves room for various interpretations~\cite{conference:Renzel2012} and different design decisions when implementing Web applications~\cite{article:Rodriguez2016}.
With the popularity of RESTful services in industry~\cite{Bogner2019-ICSA,Schermann2016} and the rapid growth of publicly available Web applications and APIs over the last two decades (see e.g. the number of newly registered APIs per year at \textit{ProgrammableWeb}\footnote{\url{https://www.programmableweb.com}}), the variety of different \enquote{REST} designs were growing as well.

The popularity, diversity, and growing complexity of Web APIs make their quality evaluation and assurance very important.
This is often manual work based on design rules or best practices, e.g. derived from the original REST constraints, and several works proposed guidance for developers when designing and implementing Web APIs~\cite{book:RestApiDesignRulebook2011,article:Palma2017,conference:Petrillo2016}.
At the same time, multiple studies analyzed the degree of REST compliance among existing real-world APIs by systematically comparing them against interface characteristics deduced from REST best practices and constraints~\cite{Neumann2018,conference:Renzel2012,article:Rodriguez2016}.
Many of these studies concluded that only a small number of existing Web APIs is indeed \textit{RESTful}, which means that the majority of Web APIs do not satisfy all required constraints.
Nevertheless, many of the APIs classified as non REST-compliant claim to be RESTful~\cite{Neumann2018}, which suggests that the term \enquote{REST} may have become a synonym for every type of API that simply offers its functionality through URIs and the HTTP~\cite{conference:Renzel2012,book:RestInPractice:2010}.

The large body of existing rules and best practices may also make it difficult for practitioners to select the most important ones.
Simultaneously, the fact that only a small number of real-world APIs seems to adhere to rules and best practices for REST-compliant APIs leads us to believe that practitioners perceive many rules aligned with Fielding's REST constraints differently in terms of their importance and impact on software quality.
We therefore see the need for an investigation from the perspective of industry practitioners.
For that purpose, we conducted a Delphi study in which we confronted eight Web API experts from industry with a catalog of REST API design rules and asked them how they rate the importance and software quality impact of each rule. This study should answer the following research questions (RQs):

\begin{enumerate}[label=\textbf{RQ\arabic*:}, leftmargin=*]
    \item Which design rules for RESTful APIs do practitioners perceive as important?
    \item How do practitioners perceive the impact of RESTful API design rules on software quality?
\end{enumerate}

These results should provide valuable input for creating an approach for the automatic rule-based quality evaluation of real-world Web APIs.
More precisely, tool support to automatically evaluate the software quality of APIs based on their conformance to important rules would assist practitioners in designing better RESTful APIs.
Based on the results, we briefly discuss requirements for such an approach and comment on the applicability of the most important rules.

The remaining paper is organized as follows:
as a basis for the selection of rules, we discuss the architectural style REST as well as existing literature about rules and best practices in Section~\ref{sec:background}.
Furthermore, we present existing Web API analysis approaches, which might serve as a foundation for an automatic rule-based quality evaluation approach.
Section~\ref{sec:study-design} explains the details of our study design.
In Section~\ref{sec:results}, we present the results of our study and discuss their interpretation and implications in Section~\ref{sec:discussion}.
Finally, we describe threats to the validity of our study (Section~\ref{sec:threats}) and close with a conclusion (Section~\ref{sec:conclusion}).

\section{Background \& Related Work}
\label{sec:background}
In this section, we discuss the principles of the architectural style REST and the reasons why this style may lead to different interpretations and design decisions. To make a selection of rules and best practices for our study, we briefly discuss existing literature presenting rules and best practices for RESTful API design. Finally, we mention existing approaches for systematic Web API analysis that might serve as a foundation for automatic rule-based quality evaluation.

\subsection{REST as an Architectural Style}
In 2000, as part of his PhD dissertation~\cite{thesis:Rest2000}, Fielding introduced REST as an architectural style that describes how distributed hypermedia systems like the World Wide Web (WWW) are designed and operated~\cite{book:RestInPractice:2010}. On the one hand, REST is a blueprint formalizing the application of existing Web techniques and standards like URIs and the HTTP~\cite{book:RestInPractice:2010}. On the other hand, it serves as a guideline for designing new applications such that they adopt these well-established Web concepts successfully and meet several quality attributes like scalability, efficiency, and reliability~\cite{article:PrincipleDesignOfTheModernWebArchitecture:2002}.  

Fielding presented REST as a framework of architectural constraints that describe the proper use of these existing Web techniques and standards and, therefore, express several design decisions for building Web applications (e.g. Web APIs). In detail, he named six core constraints that a Web application must fulfill to be RESTful, namely \textit{Client-Server}, \textit{Layered System}, \textit{Stateless Communication}, \textit{Cache}, \textit{Code-On-Demand}, and \textit{Uniform Interface}~\cite{article:PrincipleDesignOfTheModernWebArchitecture:2002}. Moreover, Fielding complemented the \textit{Uniform Interface} by four additional constraints: \textit{Identification of Resources}, \textit{Manipulation of Resources Through Representations}, \textit{Self-Descriptive Messages}, and \textit{Hypermedia as the Engine of Application State} (HATEOAS)~\cite{article:PrincipleDesignOfTheModernWebArchitecture:2002}. For a detailed descriptions of these constraints, we refer to~\cite{article:PrincipleDesignOfTheModernWebArchitecture:2002},~\cite{conference:AModelDrivenApproachForRESTCompliantServices}, and~\cite{book:RestApiDesignRulebook2011}. 

While some constraints like \textit{Client-Server} and \textit{Layered System} are automatically fulfilled because they are fundamental concepts of the WWW architecture, the concepts behind URIs and the HTTP promote other constraints like \textit{Cache} and \textit{Uniform Interface}, but cannot enforce their compliance~\cite{conference:AModelDrivenApproachForRESTCompliantServices}. In fact, a developer can ignore or misuse given mechanisms of URIs and the HTTP, which might result in a violation of these constraints. Especially in Web API design, adhering to \textit{Uniform Interface} and its four additional constraints is crucial, since this enables developers to reuse existing client code for different APIs. Individual interface styles, however, force client-side developers to implement individual clients~\cite{Neumann2018}. In our point of view, especially the \textit{Uniform Interface} constraint and its four additional constraints leave developers too much room for different interpretations, thereby leading to various interface designs in practice.

\subsection{Best Practices for REST}
\label{sec:bestPractices}
While Fielding's formal description of REST and its constraints mainly focus on the recommended behavior of a Web application and the naming of quality attributes that can be achieved by satisfying these constraints, several works examined the principles of REST. They translated them into rules and best practices to guide developers towards good RESTful design with instructions on how to satisfy them.
The existing literature in this field includes scientific articles, e.g. by Pautasso~\cite{Pautasso2014}, Petrillo et al.~\cite{conference:Petrillo2016}, and Palma et al.~\cite{article:DetectionOfRESTPatternsAndAntipatterns,article:Palma2017}, as well as textbooks, e.g. by Richardson and Ruby~\cite{book:RichardsonRuby:2007}, Massé~\cite{book:RestApiDesignRulebook2011}, and Webber et al.~\cite{book:RestInPractice:2010}. While both~\cite{book:RichardsonRuby:2007} and~\cite{book:RestInPractice:2010} proposed several best practices embedded into continuous guides for designing and implementing REST APIs, Massé~\cite{book:RestApiDesignRulebook2011} presented a catalog of 82 rules, where each rules is described concisely and is often complemented by short examples and suggestions for its implementation. Petrillo et al.~\cite{conference:Petrillo2016} compiled and presented a similar catalog, with rules inspired by the work of \cite{book:RestApiDesignRulebook2011}, but also incorporating rules and best practices from \cite{book:RichardsonRuby:2007} and other sources. 
Unsurprisingly, most of the rules and best practices in the mentioned literature focus on \textit{Uniform Interface} and its four additional constraints.

In addition to rules and best practices for REST API design, Leonard Richardson developed a maturity model~\cite{web:MaturityModel} that allows one to estimate the degree of REST compliance of a Web API. More precisely, the model consists of four levels of maturity that incorporate the principles of REST:
\begin{itemize}
\item Web APIs complying with level 0 provide their functionality over a single endpoint (URI). Moreover, they use the HTTP solely as a transport protocol for tunneling requests through this endpoint by using \texttt{HTTP POST} and without leveraging any other Web mechanisms. SOAP and XML-RPC services typically rely on level 0.
\item At level 1, Web APIs use the concept of resources, which means that they expose different endpoints (URIs) for different resources. However, at level 1, operations are still identified via URIs or specified in the request payload rather than using different HTTP methods for different operations.
\item Web APIs at level 2 use HTTP mechanisms and semantics including different verbs for different operations as well as the interpretation of status codes. Level 2 is partially aligned with the \textit{Uniform Interface} constraint of REST.
\item Level 3 Web APIs embed hypermedia controls into responses to advertise semantic relations between resources and to guide a client through the application. This final level of the maturity model addresses the HATEOAS constraint of REST. 
\end{itemize}

\subsection{Studies for the Evaluation of Web APIs}

Several publications analyzed the quality or conformity of existing RESTful APIs or proposed automatic approaches for their evaluation.

In~\cite{conference:Renzel2012}, Renzel et al. analyzed the 20 most popular RESTful Web services listed in~\textit{ProgrammableWeb} in May 2011. For that purpose, they created a catalog of 17 metrics, which are based on the REST architecture style, relevant features of the HTTP, as well as characteristics and best practices of REST as described by~\cite{book:RichardsonRuby:2007}. This catalog contains metrics like the number of different HTTP methods used in an API or whether links are embedded into representations or not. For each of these 20 APIs, they manually read the API service description to capture the values of these metrics. The analysis revealed that nearly none of the 20 Web services were indeed RESTful (only four Web services were HATEOAS-compliant, for instance). Moreover, the study identified differences in terms of adaption and interpretation of REST principles among the analyzed Web services.

Palma et al. introduced an approach for the automatic detection of REST patterns and antipatterns called \textit{SODA-R}~\cite{article:DetectionOfRESTPatternsAndAntipatterns}. They identified five patterns and eight antipatterns for REST and defined heuristics for detecting them. Palma et al. translated these heuristics into algorithms that automatically detect these (anti-)patterns on a set of APIs by invoking their interfaces. The tests against 12 real-world REST APIs revealed that developers tend to use customized headers negatively affecting the \textit{Uniform Interface}, in particular, the \textit{Self-Descriptive Messages} constraint. Moreover, eight out of 12 APIs did not embed hyperlinks into response payloads, thereby violating HATEOAS.

Rodríguez et al.~\cite{article:Rodriguez2016} analyzed more than 78 GB of HTTP network traffic to gain insights into the use of best practices when designing Web APIs. Since the analyzed data contained any kind of Web-based traffic, they used API-specific heuristics to extract REST API related requests and responses and identified the relevant APIs based on this data. Then, they validated the extracted requests and responses against 18 heuristics, which they deduced from design principles and best practices introduced in~\cite{book:RestApiDesignRulebook2011,Palma2015a,Pautasso2014}. Moreover, they mapped the heuristics to the levels of the Richardson maturity model~\cite{web:MaturityModel} to estimate the level of compliance of the identified APIs. According to the paper, only a few APIs reached level 3 of the maturity model (embedded hyperlinks into resource representations), while the majority of APIs complied with level 2.  

Petrillo et al. of~\cite{conference:Petrillo2016} compiled a catalog of 73 best practices, which we already mentioned in Section~\ref{sec:bestPractices}. Based on this catalog, they manually analyzed the documentations of three cloud provider APIs and determined whether the respective cloud provider fulfilled these best practices or not. Although all three APIs fulfilled only between 56\% and 66\% of the best practices, Petrillo et al. assessed their maturity level as acceptable. 

In~\cite{article:Palma2017}, Palma et al. presented an approach for the detection of linguistic patterns and antipatterns addressing the URI structure of REST APIs and their compliance with best practices of REST. They tested the approach with 18 real-world APIs by invoking their endpoints and analyzing their documentations. Most of the analyzed APIs used appropriate resource names and did not use verbs within URI paths, which confirms the acceptance of this best practice. However, URI paths often did not convey hierarchical structures as it should be according to another best practice.

An extensive study for investigating the quality of real-world Web APIs and estimating the level of REST compliance was presented by Neumann et al. in~\cite{Neumann2018}. The authors compared the documentations of 500 APIs claiming to be RESTful against 26 features reflecting REST principles and best practices deduced from~\cite{book:RestApiDesignRulebook2011,article:Palma2017,conference:Petrillo2016,conference:Renzel2012,article:Rodriguez2016} and other works. Similar like Renzel et al.~\cite{conference:Renzel2012}, the authors identified a very heterogeneous adoption of REST. Especially the \textit{Uniform Interface} constraint was barely adopted. The study came to the result that only four APIs (0.8\%) fully complied with REST.

The aforementioned studies primarily focus on the analysis of Web API compliance with rules, features, and best practices aligned with the principles and constraints of REST. However, there are also two approaches by Haupt et al.~\cite{conference:AFrameworkForTheStructuralAnalysisOfRestApis} and Bogner et al.~\cite{Bogner2020}.
Rather than using best practices or rules, they analyzed Web APIs based on a set of interface characteristics and calculated metrics that provide insights about the general structure and quality of Web APIs. Both approaches run completely automatically based on machine-readable API documentation, which might be interesting for our rule-based quality evaluation approach. In detail, Haupt et al.~\cite{conference:AFrameworkForTheStructuralAnalysisOfRestApis} introduced a framework to provide insights about the structure of a REST API by automatically analyzing its service description. Bogner et al.~\cite{Bogner2020} proposed a modular framework called \textit{RESTful API Metric Analyzer (RAMA)} that calculates maintainability metrics from service descriptions and enables the automatic evaluation of REST APIs. Moreover, it converts the content of a service description into a canonical meta-model, which is used for the subsequent evaluation against a set of metrics.

By analyzing real-world Web APIs, the mentioned studies revealed that only a small number of Web APIs adheres to the rules and best practices of REST and, therefore, are truly RESTful. While these studies identified mismatches between theoretical concepts of REST and implementations by examining the work of developers, the reasons for these mismatches between theory and practice remain unclear. To the best of our knowledge, the opinions of developers about theoretical REST principles as well as rules and best practices for REST API design have not been investigated so far. We therefore see the need for a study from the perspective of industry practitioners: instead of analyzing their work as previous studies did, we directly want to confront them with rules and best practices for REST API design and query their perceived importance as well as the perceived impact on software quality. Furthermore, these results should enable rule-based quality evaluations of Web APIs relying not only on rules and best practices originating from theory but also incorporating opinions of practitioners to evaluate a Web API from a more practical point of view.

\section{Research Design}
\label{sec:study-design}
In this section, we describe our followed methodology.
We first present a general overview of our research process and then describe more design details of the most important phase, the conducted Delphi study with industry experts.
For transparency and replicability, we make the most important study artifacts available online.\footnote{\url{https://doi.org/10.5281/zenodo.4643906}}

\subsection{General Process}
\begin{figure}
    \centering
    \includegraphics[width=\linewidth]{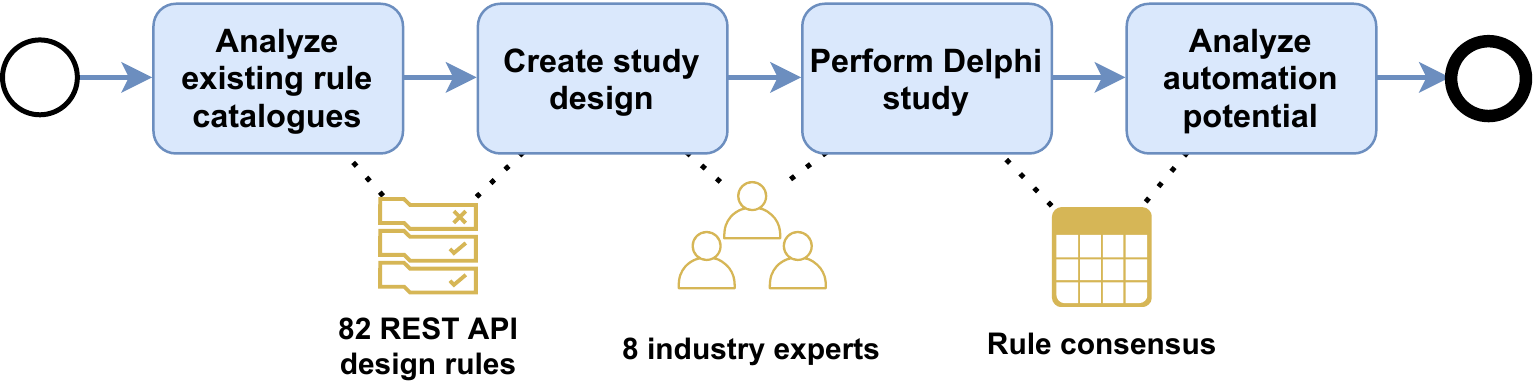}
    \caption{Overview of the research process}
    \label{fig:research-process}
\end{figure}

This research took place in several distinct stages (see Fig.~\ref{fig:research-process}).
In an initial step, we thoroughly analyzed existing rule catalogs and best practices for REST API design as well as approaches for systematic Web API analysis and related work in this area (see Section~\ref{sec:background}).
Based on the results, we selected a comprehensive list of rules (namely the catalog from Massé~\cite{book:RestApiDesignRulebook2011}) and created a detailed study protocol with the necessary materials.
We also recruited industry experts as participants and scheduled the study execution with them.
The Delphi study took place over a period of roughly nine weeks between June and August 2020.
As results, this study produced -- for each rule -- a consensus on its perceived importance (RQ1) as well as the quality attributes that experts perceived as positively impacted by adhering to it (RQ2).
Finally, we analyzed the automation potential of this consensus data, i.e. we investigated the important rules from the perspective of using them in a tool-supported approach for the rule-based quality evaluation of RESTful APIs.

\subsection{Delphi Study}
While one-time interviews or a survey could have been used to gather relevant data for RQ1 and RQ2, we instead decided to employ the Delphi method~\cite{Dalkey1963}, a structured communication approach originally designed as a forecasting technique by the RAND corporation in the 1950s.
Within several iterations, individual expert opinions and rationales for the same set of questions are collected without experts getting in direct contact with each other, thereby avoiding issues with group dynamics.
As input for the next round, participants get the anonymous results from their peers, often in aggregated form, and are encouraged to rethink their previous answers.
Iterations are stopped once sufficient consensus has been reached.
Dalkey and Helmer, two of the original authors, describe the advantages of this approach as follows~\cite{Dalkey1963}:
\enquote{The method [...] appears to be more conducive to independent thought on the part of the experts and to aid them in the gradual formation of a considered opinion.
Direct confrontation, on the other hand, all too often induces the hasty formulation of preconceived notions, an inclination to close one’s mind to novel ideas, a tendency to defend a stand once taken, or, alternatively and sometimes alternately, a predisposition to be swayed by persuasively stated opinions of others.}
Delphi studies can be conducted with different objectives and several methodological variations exist.
The method has been successfully used as a consensus technique in both information systems~\cite{Okoli2004} and software engineering~\cite{Lilja2011} research.

In our study, the objectives were to reach consensus on the importance of RESTful API design rules (RQ1) and their impact on software quality (RQ2).
For each RQ, we conducted a separate Delphi process with the same experts.

\subsubsection{Study Objects}
The objects of this study were the 82 design rules for RESTful APIs compiled by Massé~\cite{book:RestApiDesignRulebook2011}.
While several catalogs of design rules or best practices have been proposed in this area (see Section~\ref{sec:background}), Massé's collection comprises the vast majority of frequently mentioned rules in diverse areas.
It is a comprehensive collection of industrial REST API design knowledge, which has also been cited over 440 times in scientific literature\footnote{According to Google Scholar (\url{https://scholar.google.com}) in March, 2021}.
Massé organized the 82 rules nicely into five categories, namely \textit{Interaction Design with HTTP} (29), \textit{Identifier Design with URIs} (16), \textit{Metadata Design} (15), \textit{Representation Design} (13), and \textit{Client Concerns} (9), and illustrated them with helpful examples and implementation suggestions.
This pragmatic presentation makes the rules very actionable for practitioners, but also well suited for our Delphi study.
Massé relied on RFC 2119\footnote{\url{https://tools.ietf.org/html/rfc2119}} to incorporate a level of rule importance, in particular the three modifiers \texttt{may}, \texttt{should}, and \texttt{must}, optionally combined with \texttt{not}.
For example, a rule from \textit{Interaction Design with HTTP} prescribes that \enquote{22: \texttt{POST} must be used to create a new resource in a collection}.
Another example from \textit{Representation Design} states that \enquote{73: A consistent form should be used to represent error responses}.

\subsubsection{Study Participants}
For our study participants, we set out the following requirements:
(i) at least three years of experience with RESTful APIs, (ii) industry practitioners with a technical background, and (iii) from several different companies or at least divisions.
This ensured that participants had sufficient practical experience and came from different backgrounds to avoid a single company-specific perspective.
We used convenience sampling to recruit participants via our personal industry network and further referrals.
While this may introduce some bias, it also ensured that we could rely on the participants' expertise and time commitment to the study.
In the end, we were able to recruit eight participants (Table~\ref{table:participants}) with REST-related professional experience between six and 15 years (median: 11.5 years).
They were from four different global IT enterprises, where they worked in seven different divisions.
All participants were located in Germany.

\begin{table}
    \centering
	\caption{Delphi study participant demographics (PID: participant ID, CID: company ID, DID: company division ID)}
	\label{table:participants}
	\begin{tabular}{llllrr}
		PID & CID & Domain & \# of employees & DID & REST exp. (years)\\
		\hline
		\hline
		1 & \multirow{4}{*}{C1} & \multirow{4}{*}{Software \& IT Services} & \multirow{4}{*}{100k - 150k} & D1 & 6\\
        2 & & & & D2 & 15\\
        3 & & & & D2 & 7\\
        4 & & & & D3 & 13\\
        \hline
        5 & \multirow{2}{*}{C2} & \multirow{2}{*}{Software \& IT Services} & \multirow{2}{*}{50k - 100k} & D4 & 6\\
        6 & & & & D5 & 11\\
        \hline
        7 & C3 & Telecommunications & 200k - 250k & D6 & 15\\
        \hline
        8 & C4 & Technology \& Manufact. & 250k - 300k & D7 & 12\\
		\hline
	\end{tabular}
\end{table}

\subsubsection{Study Materials and Measurements}
Several artifacts were used in this study.
To get familiar with the rules and as a source for more details on specific rules, all participants received a digital copy of Massé's \textit{REST API Design Rulebook}.
Additionally, we extracted the rules and their description.
We used this content to create Excel spreadsheets to gather participants' input.
The spreadsheets also contained concrete instructions for participants and a form to gather their demographics.
For RQ1, the spreadsheet contained a list of all 82 rules with a field to rate the perceived importance of each rule.
Participants had to use a 3-point ordinal scale with the labels \texttt{low}, \texttt{medium}, and \texttt{high}.
For RQ2, a similar spreadsheet was used, but this time with fields to associate a rule with positive impact on several different quality attributes (QAs).
We used the eight top-level attributes from ISO 25010~\cite{ISO25010}, i.e. functional suitability, performance efficiency, compatibility, usability, reliability, security, maintainability, and portability.
To ensure a common understanding, participants received a link to a website with descriptions about the standard\footnote{\url{https://iso25000.com/index.php/en/iso-25000-standards/iso-25010}}.
For each rule, participants had to select which QAs where positively impacted by adhering to it.
The association was binary, i.e. a rule could either affect a specific QA (1) or not (0).
Multi-selection was possible, i.e. a rule could affect several QAs.
Both spreadsheets also contained tabs to provide the aggregated responses of the previous iteration for each rule, i.e. the number of experts who voted for the different levels of importance or the different associated QAs respectively.

Lastly, we created an analysis spreadsheet to summarize the responses and calculate \textit{consensus} and \textit{stability}.
Since required percentage agreements of 70\% and greater are common in Delphi studies~\cite{Vernon2009}, we defined 87.5\% as consensus condition for a rule, i.e. only one of the eight experts could rate it differently than the rest.
For the impact on software quality, agreement therefore meant that the exact same QAs had to be selected by seven experts.
Additionally, we considered a stopping condition based on rating stability to avoid being deadlocked in iterations without significant change of votes~\cite{Holey2007}.
As it is commonly used in Delphi studies, we chose Cohen's kappa coefficient ($\kappa$) to calculate the group stability between two iterations~\cite{Landis1977}.
Since the ratings for rule importance were on an ordinal scale, we used a weighted kappa variant~\cite{Holey2007}, i.e. the weight for changing your vote from \texttt{low} to \texttt{high} or vice versa was 2, while e.g. changing from \texttt{medium} to \texttt{high} only had a weight of 1.
We used the standard variant for the impact on software quality.
Landis and Koch~\cite{Landis1977} label $\kappa$-values between 0.81 and 1.00 as \enquote{almost perfect} agreement.
While we did not define a strict stopping threshold and the stability has to be interpreted in conjunction with the percentage agreement, $\kappa$-values of 0.90 and greater for at least two iterations indicate that it may be advisable to stop the process.

\subsubsection{Study Execution}
\begin{figure}[H]
    \centering
    \includegraphics[width=0.75\linewidth]{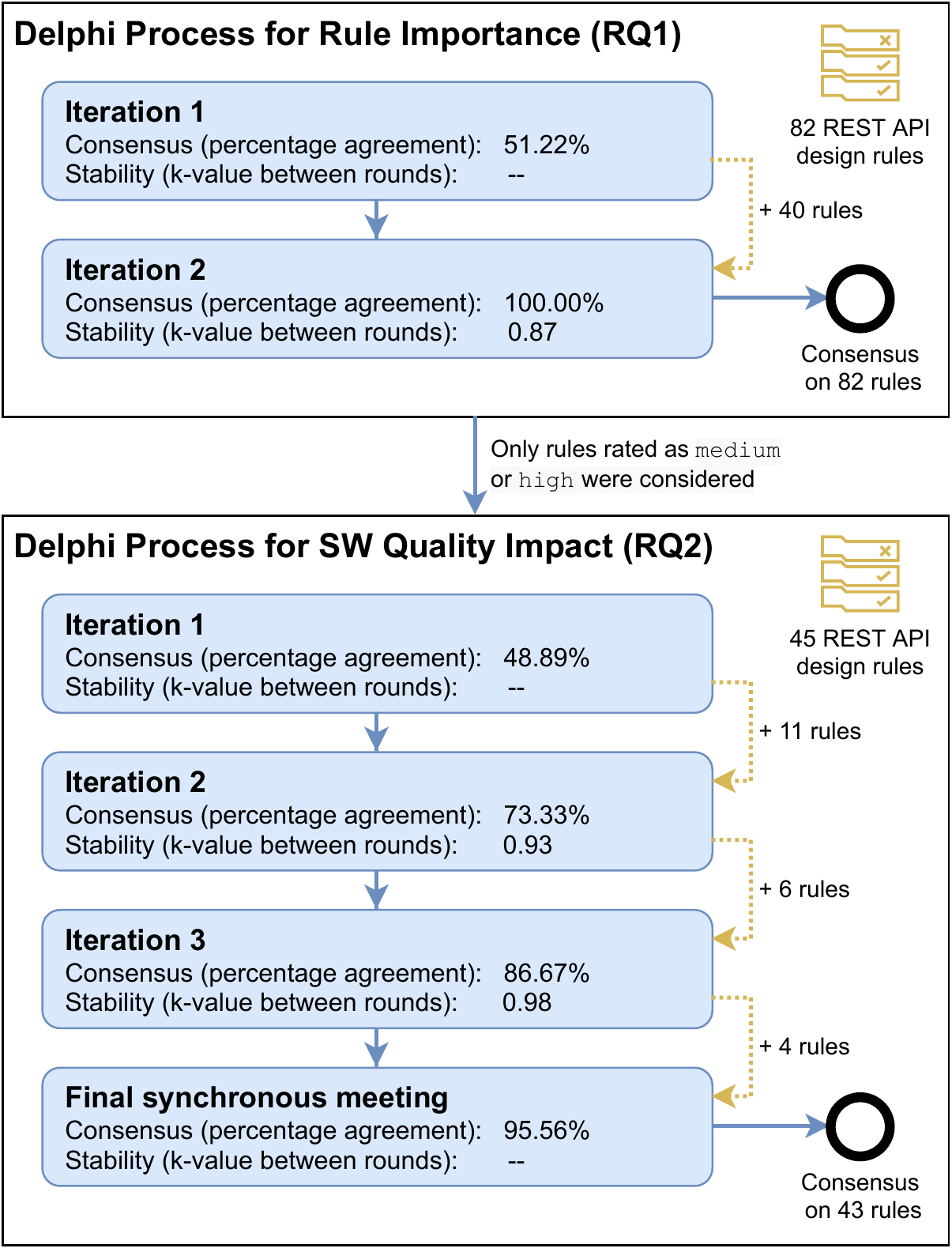}
    \caption{Overview of the Delphi study}
    \label{fig:delphi-study}
\end{figure}

Before starting the regular Delphi iterations, we scheduled personal meetings with participants to explain study procedure and provided materials.
Furthermore, we addressed open questions and communicated the response timelines.
We also encouraged participants to contact the study lead in case of uncertainties about the procedure or material during the iterations.
After these meetings, the two consecutive Delphi processes were conducted (see Fig.~\ref{fig:delphi-study}).

The goal of the first Delphi process was to reach consensus on the perceived importance of the 82 REST API design rules (RQ1).
After the first iteration, participants were in agreement about 42 of the 82 rules (51.22\%). To start the second iteration, we provided them with the aggregated results from their peers, i.e. the number of times each level of importance had been chosen for a rule.
Based on this feedback, participants were already able to reach full consensus on the 82 rules after two iterations, i.e. $\kappa$-based stopping was not necessary.

The goal of the second Delphi process was to reach consensus on the perceived positive rule impact on different software QAs (RQ2).
To reduce the cognitive load and required effort, we only advanced the 45 rules for which the experts had rated the importance as \texttt{medium} or \texttt{high} to the final stage.
Despite the substantially greater rating complexity per rule (choosing 0 or 1 for eight different QAs instead of three different importance levels), experts reached a similar percentage agreement in the first iteration, with consensus on 48.89\% of the rules (22 of 45).
Based on the provided peer responses, consensus on an additional 11 rules could be reached in iteration \#2 (73.33\% agreement with a $\kappa$-value of 0.93).
However, progress slowed down in the third iteration, where agreement on only six more rules could be reached (86.67\% agreement, $\kappa = 0.98$).
Since $\kappa$-values had been greater than 0.90 for two consecutive rounds and only six rules without consensus were left, we decided to end the iterations and instead hold a final synchronous meeting to discuss the remaining rules.
While this breaks participant anonymity and might reintroduce group dynamics bias, such meetings can lead to efficient consensus finalization.
They are often suggested after the third or even later iterations.
All participants agreed to this virtual meeting, which took place over one hour via Microsoft Teams.
During the meeting, the study lead first introduced the remaining open topics and the current state of ratings for them.
Afterwards, he initiated an open discussion.
As a result, consensus could be reached for four of the six rules.
The two rules without consensus are \enquote{54: Caching should be encouraged} and \enquote{61: JSON should be supported for resource representation}.
In both cases, five experts perceived a positive impact on functional suitability, while three experts could not be convinced by this.

\section{Study Results}
\label{sec:results}
In this section, we present the aggregated results of the conducted Delphi study, i.e. how important our eight experts judged the rules and how they perceived the rule impact on software quality.

\subsection{Perceived Importance of API Design Rules (RQ1)}
Of the 82 REST API design rules studied in the first Delphi process, participants rated 37 with \texttt{low} importance (45\%), 17 with \texttt{medium} importance (21\%), and 28 rules with \texttt{high} importance (34\%).
In other words, our experts perceived nearly half of the rules as not really important and only one third as very important.
Especially rules from the smaller categories \textit{Client Concerns} and \textit{Metadata Design} were rated as less relevant.
Most important categories were \textit{Identifier Design with URIs} (50\% of rules rated as \texttt{high}) and \textit{Interaction Design with HTTP} (45\% of rules rated as \texttt{high}), which together also make up 55\% of the total rules.
These findings are summarized in Table~\ref{table:category-importance}.
Table~\ref{table:rulesMedium} lists all rules with \texttt{medium} importance and Table~\ref{table:rulesHigh} the ones with \texttt{high} importance.

\begin{table}
    \centering
	\caption{Perceived rule importance by category (RQ1), percentages represent the fractions of total rules and fractions within the category respectively}
	\label{table:category-importance}
	\begin{tabular}{lrlr}
		Category & Total \# of Rules & Importance & \# of Rules\\
		\hline
		\hline
		\multirow{3}{*}{Identifier Design with URIs} & \multirow{3}{*}{16 (20\%)} & High & 8 (50\%)\\
		 & & Medium & 5 (31\%)\\
		 & & Low & 3 (19\%)\\
		\hline
		\multirow{3}{*}{Interaction Design with HTTP} & \multirow{3}{*}{29 (35\%)} & High & 13 (45\%) \\
		 & & Medium & 5 (17\%)\\
		 & & Low & 11 (38\%)\\
		\hline
		\multirow{3}{*}{Representation Design} & \multirow{3}{*}{13 (16\%)} & High & 4 (31\%)\\
		 & & Medium & 2 (15\%)\\
		 & & Low & 7 (54\%)\\
		\hline
		\multirow{3}{*}{Metadata Design} & \multirow{3}{*}{15 (18\%)} & High & 2 (13\%)\\
		 & & Medium & 3 (20\%)\\
		 & & Low & 10 (67\%)\\
		\hline
		\multirow{3}{*}{Client Concerns} & \multirow{3}{*}{9 (11\%)} & High & 1 (11\%)\\
		 & & Medium & 2 (22\%)\\
		 & & Low & 6 (67\%)\\
		\hline
	\end{tabular}
\end{table}

\subsection{Perceived Rule Impact on Software Quality (RQ2)}
During the second Delphi process, our experts rated the impact of the 45 rules with \texttt{medium} or \texttt{high} importance on the eight software QAs of ISO 25010.
As a result, 10 rules were assigned to one QA, 12 rules to two QAs, 18 rules to three QAs, and the remaining five rules to four different QAs, i.e. a bit more than half of the rules were perceived with a positive impact on three or more QAs.
Most associated attributes were \textit{usability} (35), \textit{maintainability} (35), and \textit{compatibility} (26).
In many cases, these QAs were impacted by the same rules.
Of the 35 rules linked to \textit{usability}, 21 were also selected for \textit{maintainability} and \textit{compatibility}.
At the same time, \textit{functional suitability} was only perceived as impacted by seven rules.
The remaining attributes were very rarely selected, with \textit{performance efficiency} being associated with three rules, \textit{reliability} and \textit{portability} with one rule each, and \textit{security} with zero rules.

\begin{figure}
    \centering
    \includegraphics[width=\linewidth]{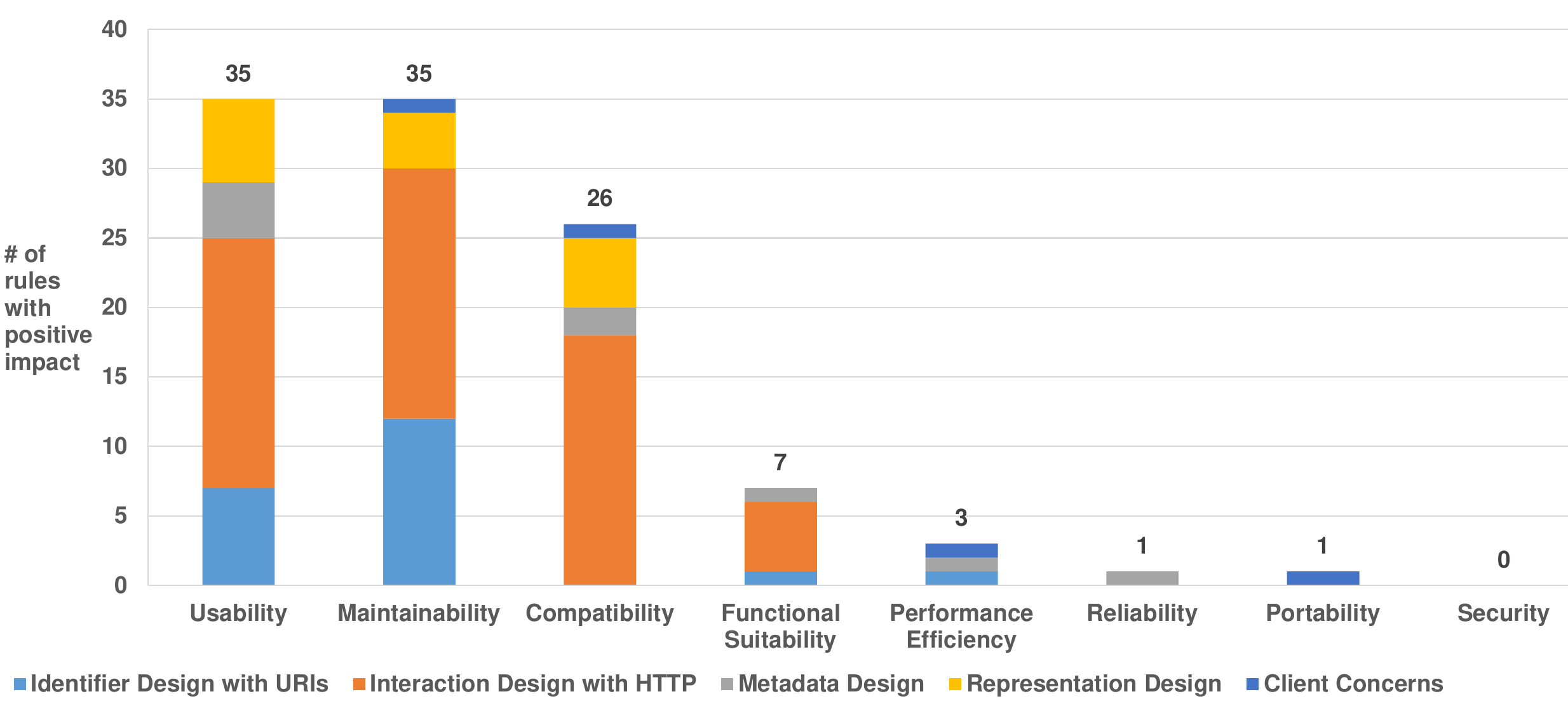}
    \caption{Perceived rule impact on software quality (RQ2), stacked by rule category}
    \label{fig:quality-impact}
\end{figure}

Fig.~\ref{fig:quality-impact} summarizes these findings and also includes the associated rule categories.
Our experts especially perceived rules from \textit{Interaction Design with HTTP} with a broad impact on software quality: all 18 rules from this category were associated with three or more QAs, with only five rules from other categories having a similar spread.
As the second largest category, \textit{Identifier Design with URIs} (13) was primarily associated with \textit{maintainability} (12) and \textit{usability} (7).
Furthermore, \textit{Representation Design} (6) spread fairly evenly among the three most impacted QAs \textit{usability} (6), \textit{compatibility} (5), and \textit{maintainability} (4).
Lastly, the rules from \textit{Metadata Design} (5) were the only ones not associated with \textit{maintainability} at all.

\begin{table}
    \centering
	\caption{All 17 REST API design rules perceived with \texttt{medium} importance (RQ1)}
	\label{table:rulesMedium}
	\begin{tabular}{p{0.04\textwidth}>{\raggedright}p{0.1\textwidth}>{\raggedright\arraybackslash}p{0.76\textwidth}}
		
ID & Category                    & Rule                                                                 \\
\hline
\hline
2  & URIs      & A trailing forward slash (/) should not be included in URIs                        \\
6  & URIs      & File extensions should not be included in URIs                                     \\
11 & URIs      & A plural noun should be used for store names                                       \\
12 & URIs      & A verb or verb phrase should be used for controller names                          \\
15 & URIs      & The query component of a URI may be used to filter collections or stores           \\
20 & HTTP      & \texttt{PUT} must be used to both insert and update a stored resource                       \\
26 & HTTP      & \texttt{200} (\enquote{OK}) should be used to indicate nonspecific success                          \\
31 & HTTP      & \texttt{301} (\enquote{Moved Permanently}) should be used to relocate resources                     \\
41 & HTTP      & \texttt{406} (\enquote{Not Acceptable}) must be used when the requested media type cannot be served \\
43 & HTTP      & \texttt{412} (\enquote{Precondition Failed}) should be used to support conditional operations       \\
47 & Meta      & Content-Length should be used                                                      \\
51 & Meta      & Location must be used to specify the URI of a newly created resource               \\
54 & Meta      & Caching should be encouraged                                                       \\
63 & Repr.     & XML and other formats may optionally be used for resource representation           \\
69 & Repr.     & Minimize the number of advertised \enquote{entry point} API URIs                           \\
79 & Client    & The query component of a URI should be used to support partial response            \\
82 & Client    & CORS should be supported to provide multi-origin read/write access from JavaScript \\
\hline
	\end{tabular}
\end{table}

\begin{table}
    \centering
	\caption{All 28 REST API design rules perceived with \texttt{high} importance (RQ1)}
	\label{table:rulesHigh}
	\begin{tabular}{p{0.04\textwidth}>{\raggedright}p{0.1\textwidth}>{\raggedright\arraybackslash}p{0.76\textwidth}}
		
ID & Category                    & Rule                                                                 \\
\hline
\hline
1  & URIs      & Forward slash separator (/) must be used to indicate a hierarchical relationship                           \\
3  & URIs      & Hyphens (-) should be used to improve the readability of URIs                                              \\
4  & URIs      & Underscores (\_) should not be used in URI                                                                 \\
5  & URIs      & Lowercase letters should be preferred in URI paths                                                         \\
9  & URIs      & A singular noun should be used for document names                                                          \\
10 & URIs      & A plural noun should be used for collection names                                                          \\
13 & URIs      & Variable path segments may be substituted with identity-based values                                       \\
14 & URIs      & CRUD function names should not be used in URIs                                                             \\
17 & HTTP      & \texttt{GET} and \texttt{POST} must not be used to tunnel other request methods                                              \\
18 & HTTP      & \texttt{GET} must be used to retrieve a representation of a resource                                                \\
22 & HTTP      & \texttt{POST} must be used to create a new resource in a collection                                                 \\
23 & HTTP      & \texttt{POST} must be used to execute controllers                                                                   \\
24 & HTTP      & \texttt{DELETE} must be used to remove a resource from its parent                                                   \\
27 & HTTP      & \texttt{200} (\enquote{OK}) must not be used to communicate errors in the response body                                     \\
28 & HTTP      & \texttt{201} (\enquote{Created}) must be used to indicate successful resource creation                                      \\
30 & HTTP      & \texttt{204} (\enquote{No Content}) should be used when the response body is intentionally empty                            \\
34 & HTTP      & \texttt{304} (\enquote{Not Modified}) should be used to preserve bandwidth                                                  \\
37 & HTTP      & \texttt{401} (\enquote{Unauthorized}) must be used when there is a problem with the client’s credentials                    \\
38 & HTTP      & \texttt{403} (\enquote{Forbidden}) should be used to forbid access regardless of authorization state                        \\
44 & HTTP      & \texttt{415} (\enquote{Unsupported Media Type}) must be used when the media type of a request’s payload cannot be processed \\
45 & HTTP      & \texttt{500} (\enquote{Internal Server Error}) should be used to indicate API malfunction                                   \\
46 & Meta      & Content-Type must be used                                                                                  \\
57 & Meta      & Custom HTTP headers must not be used to change the behavior of HTTP methods                                \\
61 & Repr.     & JSON should be supported for resource representation                                                       \\
71 & Repr.     & A consistent form should be used to represent media type formats                                           \\
72 & Repr.     & A consistent form should be used to represent media type schemas                                           \\
73 & Repr.     & A consistent form should be used to represent error responses                                              \\
74 & Client    & New URIs should be used to introduce new concepts \\
\hline
	\end{tabular}
\end{table}

\section{Discussion}
\label{sec:discussion}
The interpretation of the study results leads to several findings with interesting potential implications for research and practice.

Concerning the importance of rules, we identified substantial differences between Massé's provided categorization and our participants' verdict.
In total, Massé classified 10 of the 82 rules with the lowest level \texttt{may}, 50 with \texttt{should}, and 22 with \texttt{must}.
Conversely, our participants perceived 45\% of proposed rules (37) to be of \texttt{low} importance, among them seven classified as \texttt{must} and 24 as \texttt{should} by Massé.
Additionally, they only selected \texttt{medium} for 17 rules, but \texttt{high} importance for 28 rules.
Among the latter, Massé classified 15 with \texttt{should} and even one with \texttt{may} (\enquote{13: Variable path segments may be substituted with identity-based values}).
All in all, the verdict of our industry experts was more \enquote{extreme}.
Despite the intuitive expectation that such a consensus technique may lead to a more balanced compromise, participants perceived many more rules as not really important and a few more as very important in comparison to Massé.
This may suggest that practitioners prefer a small set of important rules instead of a large collection in which many rules are situational.

When analyzing rules related to the Richardson maturity model~\cite{web:MaturityModel}, participants rated the vast majority of rules necessary to reach \textit{Level 2: HTTP Verbs} with \texttt{medium} or \texttt{high} importance.
This includes the rules 17-25 about the correct usage of HTTP verbs (with the exception of \texttt{HEAD}, \texttt{OPTIONS}, and interestingly \texttt{PUT}), but also the rules 26-45 about correct status code usage.
Conversely, nearly all of the rules associated with \textit{Level 3: Hypermedia Controls} were categorized with \texttt{low} importance, including the \textit{Representation Design} rules 65-68 and also the HATEOAS rule \enquote{70: Links should be used to advertise a resource’s available actions in a state-sensitive manner}.
This indicates that our experts saw maturity level 2 as strongly desirable but also as sufficient, which is consistent with the results of Rodriguez et al.~\cite{article:Rodriguez2016} and Neumann et al.~\cite{Neumann2018}.
Reaching level 3 may not be a priority for industry, even though it is one of the original RESTful constraints postulated by Fielding~\cite{thesis:Rest2000}.

Concerning the rule impact on software quality, the vast majority of rules focus on usability (from an API user perspective), maintainability, or compatibility, with many rules influencing two or even all three of these QAs.
Other QAs are less represented or even not at all.
While it makes sense that service interface design may not be related to all QAs to the same extent, especially the absence of rules with an influence on performance efficiency and security is surprising.
Intuitively, interface design should have a substantial influence on these two QAs, but the focus of Massé's rules seems to lie elsewhere.
To evaluate these other QAs, different rule catalogs or approaches are therefore necessary.

Lastly, the greatest benefit should in theory be achieved by adhering to those rules that were both rated with \texttt{high} importance and associated with several QAs at once.
In our sample, 17 rules with \texttt{high} importance impact at least three QAs, 13 of them from the category \textit{Interaction Design with HTTP}, three from \textit{Representation Design} (rules 71-73), and rule 57 from \textit{Metadata Design}.
The majority of them are again related to the correct usage of HTTP verbs and status codes, while rules 71-73 are concerned with the consistent structuring of media type formats, media type schemas, and error responses.
This again highlights the fact that, in addition to consistency, our participants perceived reaching Richardson maturity level 2 as very important.
These 17 rules are primary candidates for automatic evaluation approaches.

As part of our future work, we therefore plan to create tool support to evaluate Web APIs based on design rules.
We start with the rules that practitioners assessed with \texttt{high} importance.
This rule-based evaluation tool should assess the quality of a Web API by reporting detected rule violations and calculating individual scores for the QAs linked to these rules.
Abstraction and modularity should guarantee that rules can easily be added, altered, or removed, e.g. according to new empirical data or based on company or team preferences.
Reported rule violations should also be actionable for practitioners, e.g. by reporting the concrete location of findings and potential ways to address the violations.
The SODA-R approach by Palma et al.~\cite{article:DetectionOfRESTPatternsAndAntipatterns} may serve as inspiration for our tool from both a methodological as well as a technical point of view.
Similar as with SODA-R, for each rule, we will first identify REST properties and then define heuristics that indicate whether the analyzed API satisfies the respective rule.
These properties can be either static and obtainable from REST API documentation (e.g. the set of URI templates indicating whether an API contains underscores (\texttt{\_}), see rule 4) or dynamic and can be inferred from response data at runtime (e.g. a response status code). 
In future work, we will translate these heuristics into algorithms that automatically verify rule compliance of Web APIs.
The authors of the SODA-R approach introduced heuristics for 13 REST (anti-)patterns~\cite{article:DetectionOfRESTPatternsAndAntipatterns} that partially overlap with some of our studied rules and, therefore, might be adaptable to our tool (e.g. heuristics for rule 26 and 46). Further heuristics applicable to the rules 4, 5, and 14 are defined by Rodríguez et al. in~\cite{article:Rodriguez2016}. Moreover, in a second work by Palma et al.~\cite{article:Palma2017}, further approaches for detecting violations of the rules 1, 4, 5, 9, 10, and 14 are presented that might be incorporated in our tool. 

\section{Threats to Validity}
\label{sec:threats}
Several threats to the validity of this study have to be considered, which we discuss according to dimensions proposed by Wohlin et al.~\cite{Wohlin2003}.

\textit{Construct validity} describes the relation between the studied high-level concepts and the operationalized metrics to measure them, in other words, whether we actually measured what we wanted to measure.
We studied the importance and software quality impact of design rules and measured how our Delphi participants perceived these concepts, i.e. we collected expert opinions.
While experts often have solid experience-based judgment, they can also be wrong, even collectively.
As long as our results are clearly treated as expert opinions, this is not an issue, especially since industry perception is important for research in this field.
However, once we start to implement some of these rules, we will need more empirical data to judge their effectiveness, preferably non-opinion-based data.

\textit{Internal validity} is concerned with the applied empirical rigor and potentially hidden factors that may have influenced the results.
While we diligently designed our research protocol, adhered to it during the study, and relied on an established consensus technique, it is possible that confounding factors could have partially influenced the results.
One such factor could be the misinterpretation of a rule or QA.
To mitigate this, all participants received materials describing these concepts, and we encouraged them to ask clarifying questions.
We perceive the risk of this threat as very small.
A more substantial threat could be that participants knew the original rule importance level, as Massé encoded it directly into the rule text (\texttt{may}, \texttt{should}, \texttt{must}).
This could have lead to an anchoring effect, thereby biasing participants in their own verdict.
However, as shown above, our participants were not afraid to adjust the initial importance level, leading to substantial differences between their and Massé's ratings.
Since rewriting the rule texts could have introduced even greater inconsistencies, we still perceive this threat as fairly low.
A last threat in this area could be the final synchronous meeting in our Delphi study.
By breaking anonymity and facilitating synchronous exchange, we opened the door for potential influences related to group dynamics.
Nonetheless, this was only for six of the 82 rules and only related to the QAs, not the importance.
Moreover, no complete consensus was reached in this discussion, thereby reducing the probability that most participants yielded to the reputation or behavior of a few.

Lastly, \textit{external validity} refers to the generalization potential of results, i.e. how well the results obtained in our specific setting carry over to different contexts.
Only the 82 rules from Massé's catalog were considered in this study.
While they provide a comprehensive collection of REST API design knowledge and substantially overlap with other catalogs, they are not the only available source and were published in 2011, i.e. roughly 10 years ago.
Moreover, our software quality analysis revealed a focus on usability, maintainability, and compatibility.
Rules with different focus and level of abstraction should be considered in future research.
Additionally, only eight experts participated in our Delphi study.
While they reached strong consensus and were from different companies and divisions, their general contexts were still fairly similar (large IT enterprises, enterprise application development, located in Germany).
All in all, it is possible that experts with very different background would come to a different verdict, e.g. when considering the importance of HATEOAS-related rules.

\section{Conclusion}
\label{sec:conclusion}
In this paper, we presented the results of a Delphi study with the goal to analyze the perceived importance (RQ1) and positive software quality impact (RQ2) of REST API design rules from the perspective of industry practitioners.
As study objects, we selected the catalog of 82 REST API design rules compiled by Massé.
As participants, we recruited eight industry experts from four different IT companies with between six and 15 years of experiences with REST.
As consensus results, our participants rated 37 rules with \texttt{low} importance (45\%), 17 with \texttt{medium} importance (21\%), and 28 rules with \texttt{high} importance (34\%).
From the 45 rules with \texttt{medium} or \texttt{high} importance, the quality attributes impacted the most were usability (35), maintainability (35), and compatibility (26).
Other attributes from ISO 25010 were less impacted or not at all.
Overall, the experts rated the rule importance differently than Massé, choosing the lowest level for many more rules and the highest level for a few more.
They also perceived rules for reaching level 2 of the Richardson maturity model as very important, but not the ones for level 3.
The acquired consensus data may serve as valuable input for designing a tool-supported approach for the automatic quality evaluation of RESTful APIs.
Follow-up research should carefully analyze the automation potential of these rules, e.g. for static or dynamic analysis, as well as how to report and aggregate the findings.
For finally creating such rule-based tool support, existing approaches may serve as a methodological or technical basis.
To support such endeavors, we make our study artifacts available online.\footnote{\url{https://doi.org/10.5281/zenodo.4643906}}

\subsubsection*{Acknowledgments}
We kindly thank Tobias Hallmayer for his assistance with study planning, execution, and analysis.
This research was partially funded by the Ministry of Science of Baden-Württemberg, Germany, for the doctoral program \textit{Services Computing} (\url{https://www.services-computing.de/?lang=en}).

%
%
\bibliographystyle{splncs04}
\bibliography{sources}

\end{document}